\documentclass[12pt]{article}
\usepackage{epsfig}
\def\gsim{\;\raise0.3ex\hbox{$>$\kern-0.75em\raise-1.1ex\hbox{$\sim$}}\;}
\def\lsim{\;\raise0.3ex\hbox{$<$\kern-0.75em\raise-1.1ex\hbox{$\sim$}}\;}
\newcommand{\be}{\begin{equation}}

\newcommand{\ee}{\end{equation}}
\newcommand{\bea}{\begin{eqnarray}}
\newcommand{\eea}{\end{eqnarray}}
\newcommand{\bt}{\begin{tabular}}
\newcommand{\et}{\end{tabular}}
\newcommand{\ba}{\begin{array}}
\newcommand{\ea}{\end{array}}

\begin{document}
\setlength{\unitlength}{1mm}

\setlength{\unitlength}{1mm} {\hfill
    $\ba{r}
    \mbox{DSF 23/2004} \\
    \mbox{astro-ph/0407638}
    \ea$}\vspace*{1cm}

\begin{center}
{\Large \bf Earth--skimming UHE Tau Neutrinos at the Fluorescence
Detector of Pierre Auger Observatory}
\end{center}

\bigskip\bigskip

\begin{center}
{\bf C. Aramo}$^{1}$, {\bf A. Insolia}$^{2}$, {\bf A.
Leonardi}$^{2}$, {\bf G. Miele}$^{1}$, {\bf L. Perrone}$^{3}$,
{\bf O. Pisanti}$^{1}$, {\bf D.V. Semikoz}$^{4,5}$
\end{center}
\vspace{.5cm}

\noindent {\it $^{1}$ Dipartimento di Scienze Fisiche, Universit\`{a} di Napoli ``Federico II'' and  Istituto
Nazionale di Fisica Nucleare Sezione di Napoli, Complesso Universitario di
Monte S. Angelo, Via Cinthia, I-80126 Napoli, Italy.\\ $^{2}$ Dipartimento
di Fisica e Astronomia, Universit\`{a} di Catania and Istituto Nazionale di
Fisica Nucleare Sezione di Catania, Via S. Sofia 64, I-95123 Catania,
Italy. \\ $^{3}$ Fachbereich C, Sektion Physik, Universit\"{a}t Wuppertal,
D-42097 Wuppertal, Germany. \\ $^{4}$ Department of Physics and Astronomy,
UCLA, Los Angeles, CA 90095-1547 USA.\\ $^{5}$ INR RAS, 60th October
Anniversary prospect 7a, 117312 Moscow, Russia.}

\bigskip\bigskip\bigskip

\begin{abstract}
Ultra high energy neutrinos are produced by the interaction of
hadronic cosmic rays  with the cosmic radiation background. More
exotic scenarios like {\it topological defects} or {\it new
hadrons} predict even larger fluxes. In particular,
Earth--skimming tau neutrinos could be detected by the
Fluorescence Detector (FD) of Pierre Auger Observatory. A detailed
evaluation of the expected number of events has been performed for
a wide class of neutrino flux models. An updated computation of
the neutrino--nucleon cross section and of the tau energy losses
has been carried out. For the most optimistic theoretical models,
about one Earth--skimming neutrino event is expected in several
years at FD.

\end{abstract}

\vspace*{2cm}

\begin{center}
{\it PACS numbers: 95.85.Ry, 13.15.+g, 96.40.Tv, 95.55.Vj, 13.35.Dx;}
\end{center}

\thispagestyle{empty}
\setcounter{page}{0}

\newpage
\baselineskip=.8cm


\section{Introduction}

The measurement of Ultra High Energy Cosmic Rays (UHECR) flux is
the goal of a wide class of past, present and future detectors
\cite{Volcano}-\cite{Euso}. UHE neutrinos are expected to be
produced by the interaction of hadronic matter with the
surrounding radiation/matter. A search for this signal is
currently performed by several Neutrino Telescopes
\cite{Baikal}-\cite{Nestor}.

Neutrinos with energy above $10^{17}$ eV are expected to originate
from the interaction of UHE cosmic rays with the Cosmic Microwave
Background (CMB) {\it via} the $\pi$-photoproduction, $p +
\gamma_{CMB} \rightarrow n + \pi^+$, the so-called {\it cosmogenic
neutrinos} \cite{cosmogenic}. The prediction for such a flux is
however affected by several uncertain physical quantities, namely
the spatial distribution of astrophysical sources, the ejected
proton fluxes (if proton) and the way of modelling the diffuse
extragalactic electromagnetic background in the different
frequency regions. One can assume a reasonable ansatz for all
these quantities combined with the measurement of the diffuse
photon flux in the GeV region by EGRET \cite{egret}, and the
AGASA/HiRes data. These models and the predictions of more exotic
scenarios have been exhaustively discussed in several papers (see
for example Ref.s \cite{Kalashev:2002kx,Semikoz:2003wv}).

High energy neutrinos are hardly detected, as they are almost
completely shadowed by Earth and rarely interact with the
atmosphere. An EeV neutrino has an interaction length of the order
of 500 km water equivalent in rock and, even crossing horizontally
the atmosphere (360 meters water equivalent), only one neutrino
out of thousand will be interacting. Due to the very low expected
flux and the small neutrino-nucleon cross section, km$^3$-neutrino
telescopes and giant surface arrays have very few chances of
detection.

In this framework, an interesting strategy for $\nu_\tau$
detection is described in Ref.s
\cite{Capelle:1998zz}--\cite{Bertou:2001vm}. As shown for example
in Figure 1 of Ref. \cite{Beacom:2001xn}, for energy between
$10^{18}$ and $10^{21}$ eV the $\tau$ decay length is not much
larger than the corresponding interaction range. Thus, an
energetic $\tau$, produced by Charged Current (CC) $\nu_\tau$
interaction not too deep under the surface of the Earth, has a
chance to emerge in the atmosphere as an upgoing particle. Unlike
$\tau$'s, muons crossing the rock rapidly loose energy and decay.
Almost horizontal $\nu_\tau$, just skimming the Earth surface,
will cross an amount of rock of the order of their interaction
length and thus will be able to produce a corresponding $\tau$,
which might shower in the atmosphere and be detected. In order to
estimate the number of upgoing $\tau$ expected in the Pierre Auger
Observatory (PAO), one needs to know the value of the
neutrino-nucleon cross section for CC interaction, $\nu_\tau \, +
\, N \rightarrow \tau \, + \, X$.

The aim of this paper is to estimate the number of possible
upgoing $\tau$ showers which the Fluorescence Detector (FD) of PAO
could detect. The predictions are analyzed with respect to their
dependence on different neutrino fluxes and by using a new
estimate of $\nu-N$ cross section.

The paper is organized as follows. In section \ref{sec:nuflux}
different models for neutrino flux predictions are discussed. In
section \ref{sec:nucross}, the general features of deep inelastic
neutrino cross sections are outlined for neutrino energies up to
10$^{21}$ eV. In section \ref{sec:tauloss}, high energy $\tau$
propagation through matter is illustrated by considering all the
relevant interaction mechanisms. Average values for $\tau$ energy
loss are provided by taking into account the recent calculations
of photonuclear interaction given in Ref. \cite{bugaev03}. In
section \ref{sec:earthskimming} the number of expected upgoing
$\tau$ showers is derived and discussed. Finally in section
\ref{sec:conclusions}, we give our conclusions and remarks.

\section{Neutrino flux estimates}
\label{sec:nuflux}

Ultra High Energy protons, with energy above $\sim 10^{20}$ eV,
travelling through the universe mostly loose their energy {\it
via} the interaction with CMB radiation. The large amount of
charged and neutral pions produced will eventually decay in
charged leptons, neutrinos ({\it cosmogenic neutrinos}
\cite{cosmogenic}), and high energy gamma rays.

\begin{figure}[h]
\begin{center}
\epsfig{figure=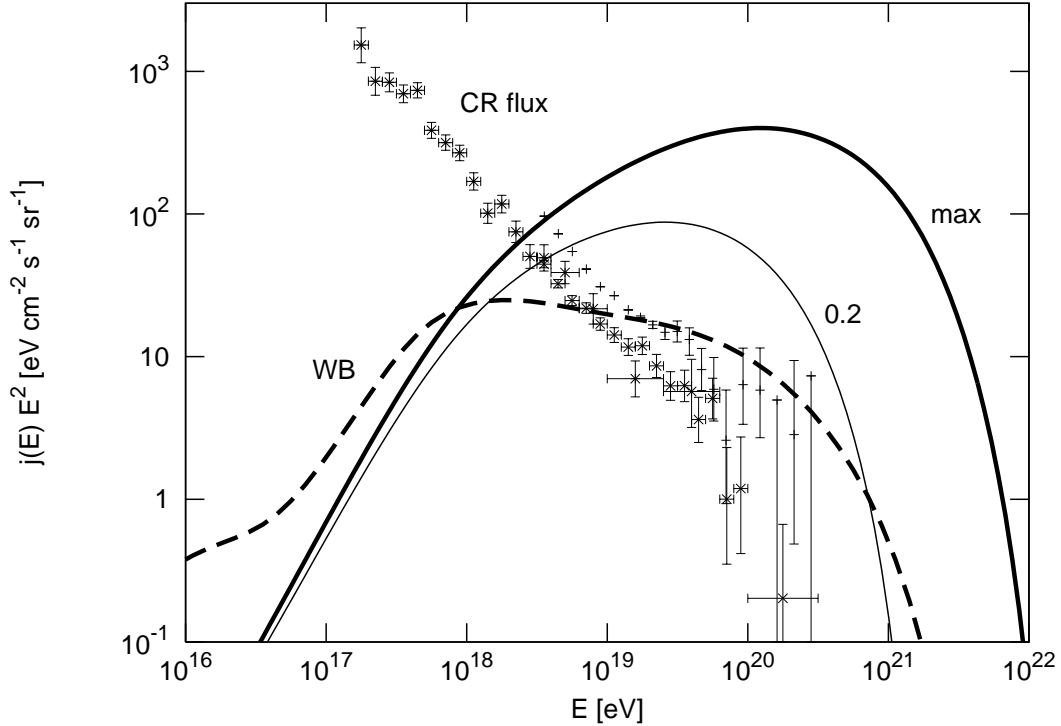,height=10cm} \caption{Cosmogenic
neutrino fluxes as a function of energy. Thick solid line is for
an initial proton flux $\propto 1/E$, by assuming that the EGRET
flux is entirely due to $\pi$-photoproduction (GZK-H). Thin solid
line shows the neutrino flux when the associated photons
contribute only up to 20\% in the EGRET flux (GZK-L). Dashed line
stands for an initial proton flux $\propto 1/E^2$ (GZK-WB). The
experimental points represent the UHECR flux measured by AGASA
(crosses) and HiRes (stars).} \label{nu_gzk}
\end{center}
\end{figure}
At the GeV energy range the extragalactic diffuse gamma-ray
background was measured by the EGRET experiment \cite{egret}. This
measurement provides an upper bound for possible neutrino fluxes
from pion production. In particular, it gives the expected maximum
flux of cosmogenic neutrinos from an initial spectrum of measured
UHE protons \cite{berez2}. It is worth noticing that, since at
least part of UHECR are protons, the existence of cosmogenic
neutrinos is guaranteed, even if their flux is very uncertain. In
Figure \ref{nu_gzk} the GZK neutrino flux for three possible
scenarios is plotted. The thick solid line gives the case of an
initial proton flux $\propto 1/E$, by assuming in addition that
the EGRET flux is entirely due to $\pi$-photoproduction (GZK-H).
The thin solid line shows the neutrino flux when the associated
photons contribute only up to 20\% in the EGRET flux (GZK-L). The
dashed line stands for the conservative scenario of an initial
proton flux $\propto 1/E^2$ (GZK-WB). In this case the neutrino
flux is compatible with the so--called Waxman-Bahcall limit
\cite{wb}. Note that no lower bounds can be set for the cosmogenic
neutrino flux. In particular, in the most conservative but rather
unrealistic case, the astrophysical sources cannot accelerate
protons up to energies above GZK cutoff, and thus the secondary
neutrinos will be produced in negligible quantities. All neutrino
fluxes presented in this section were calculated by a propagation
code \cite{code2} which takes into account the neutrino production
via $\pi$-photoproduction  with microwave, infrared, optical and
radio photon backgrounds, as well as in neutron decay.
\begin{figure}[h]
\begin{center}
\epsfig{figure=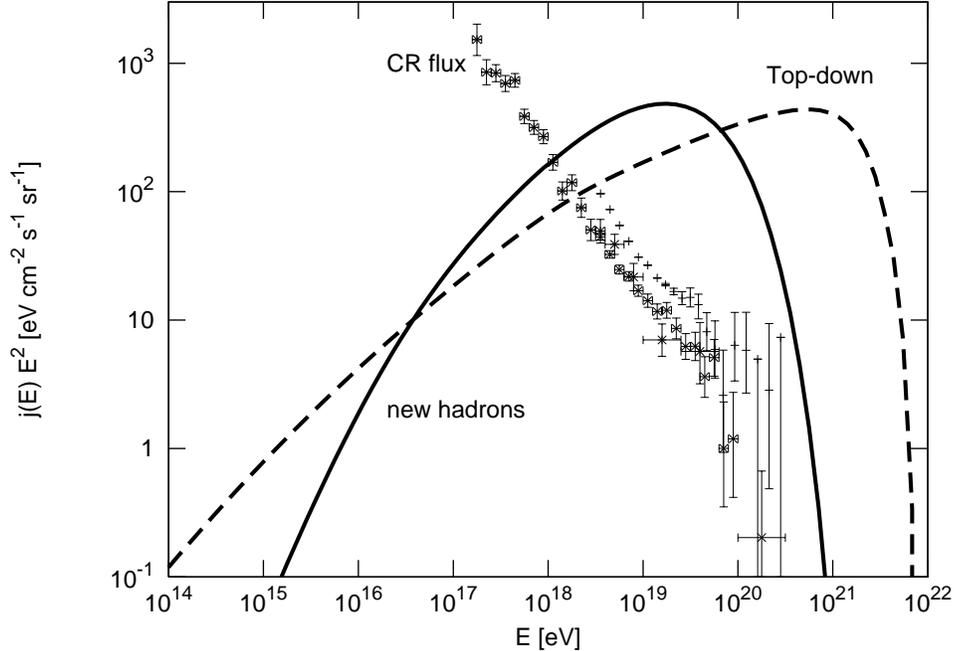,height=9cm} \caption{Neutrino
fluxes in exotic UHECR models. Solid line represents the neutrino
flux prediction in a model with {\it new hadrons} (NH)
\cite{hadrons}, whereas the dashed line is the neutrino flux for a
topological defect model (TD) \cite{sigl_review}.}
\label{nu_exotic}
\end{center}
\end{figure}

Most of the models trying to explain highest energy cosmic rays
($E>10^{20}$ eV) in terms of exotic particles, predict a large
associated flux of neutrinos. In Figure \ref{nu_exotic}, the
expected neutrino flux for two of such scenarios is plotted. One
of them is the model of {\it new hadrons} (NH) \cite{hadrons},
with mass $M\sim 2-5$ GeV, capable of generating UHECR events
above GZK cutoff. In SUSY theories, for example, the new hadrons
are bound states of light bottom squarks or gluinos and, once
produced in suitable astrophysical environments, can reach the
Earth without significant energy losses. In spite the production
of new hadrons is a subdominant process, it generates a large
number of neutrinos (see Figure \ref{nu_exotic}).

The dashed line in Figure \ref{nu_exotic} shows the neutrino flux
for a Topological Defects model (TD) (for a review see
\cite{sigl_review}). In this case UHECR events with energy
$E>10^{20}$ eV are explained in terms of $\gamma$'s which are
produced in the decay of heavy particles with mass of the order of
$10^{22-23}$ eV. As in the previous case, the associated neutrino
flux for this kind of models is extremely large.

We do not discuss here the so--called Z-burst scenarios, which
attempts to explain UHECR as products of Z-boson decay. These
models are in fact strongly disfavored \cite{Semikoz:2003wv} by
the upper bounds on UHE neutrino flux put by FORTE \cite{forte}
and GLUE \cite{glue} and by the cosmological limits on neutrino
mass set by WMAP and LSS data
\cite{Hannestad:2003xv,Crotty:2004gm,Fogli:2004as}.

In the following sections we will estimate the sensitivity of PAO
to the UHE tau neutrino flux both in the case of cosmogenic
neutrinos (Figure \ref{nu_gzk}) and of exotic models (Figure
\ref{nu_exotic}).

\section{Neutrino-Nucleon cross section in the extremely high energy limit}
\label{sec:nucross}

At energy above 1 GeV neutrino-atoms interaction is dominated by
the process of {\it Deep Inelastic Scattering} (DIS) on nucleons,
since the contributions of both elastic and quasi-elastic
interactions become negligible. The effect of the neutrino
scattering with atomic electrons will not be taken into account
here, since the cross section for this process is, at each energy,
about three orders of magnitude lower than the neutrino-nucleon
cross section\footnote{The only exception is the resonant
$\bar{\nu}_{e}\rightarrow W^{-}$ production, occurring at
$E_{\bar{\nu}_{e}}= 6.3$ PeV, whose contribution to the total
event rate remains nevertheless negligible~\cite{quiggcross}.}.

Detectable leptons are produced through Charged Current
interaction,
\begin{equation}
\nu_{l} (\bar{\nu}_{l}) + N \rightarrow l^{-} (l^{+}) + X \,\,\, ,
\label{cc}
\end{equation}
whereas Neutral Current (NC) interaction causes a modulation in
the spectrum of the interacting neutrinos,
\begin{equation}
\nu_{l} (\bar{\nu}_{l}) + N \rightarrow \nu_{l} (\bar{\nu}_{l}) +
X' \,\,\, .
\label{nc}
\end{equation}
These total cross sections can be written in terms of differential
ones as follows
\begin{eqnarray}
\sigma_{CC}^{\nu N}(E_{\nu})&=&  \int_{0}^{1-\frac{m_{\l}}{E_{\nu}}}
              \frac{d\sigma_{CC}^{\nu N}}{dy} \,(E_{\nu},y) \, dy \,\,\, ,\\
\sigma_{NC}^{\nu N}(E_{\nu})&=&  \int_{0}^{1}
             \frac{d\sigma_{NC}^{\nu N}}{dy} \,(E_{\nu},y) \, dy \,\,\, ,
\end{eqnarray}
where  $E_{\nu}$ is the energy of the incoming neutrino, $m_{\l}$
is the mass of the outgoing charged lepton and $y$ is the
inelasticity parameter, defined as
\begin{equation}
y_{CC,NC}=1-\frac{E_{\l}}{E_{\nu}}\,\,\, ,
\end{equation}
with $E_{\l}$  the energy of the outgoing charged (for CC) or
neutral (for NC) lepton.

\subsection{Deep inelastic neutrino cross sections}
\label{subsec::cross}

\begin{table}[thb]
\centering
\begin{tabular}{||c|c|c||}\hline\hline
 Energy (GeV) & $<y_{CC}>$ & $<y_{NC}>$ \\ \hline\hline
  $10^{7}$    &  0.2388  & 0.2449   \\
  $10^{8}$    &  0.2180  & 0.2223  \\
  $10^{9}$    &  0.2019  & 0.2052 \\
  $10^{10}$   &  0.1900  & 0.1928 \\
  $10^{11}$   &  0.1785  & 0.1821 \\
  $10^{12}$   &  0.1542  & 0.1601 \\ \hline
\end{tabular}
\caption{Average inelasticity parameter for CC ($<y_{CC}>$) and NC
($<y_{NC}>$) interaction for different incoming neutrino energy.}
\label{table::y}
\end{table}
Tau neutrino-nucleon cross sections have been calculated following
the approach of Ref.~\cite{quiggcross}, based on the
renormalization-group-improved parton model, and by using the most
recent data on the parton structure functions of nucleons.

The cross sections can be written in terms of the Bjorken scaling
variables $y$ and $x=Q^2/2MyE_{\nu}$, where $-Q^2$ is the
invariant momentum transferred between the incoming neutrino and
the outgoing lepton. Details of nucleon structure become important
at very high-energy where available data are very poor or totally
missing. As a consequence of this, the lack of knowledge of the
parton structure functions at very low $x$ ($x\ll 10^{-5}$)
dominates the whole uncertainty on the cross section calculations
at very high-energy. In the present analysis we have used the
CTEQ6~\cite{cteq6} parton distribution functions in the DIS
factorization scheme.\\ The $Q^2$--evolution is realized by the
next-to-leading order Dokshitzer-Gribov-Lipatov-Altarelli-Parisi
equations \cite{Gribov:1972rt}--\cite{Altarelli:1977zs}. The CTEQ6
distributions are particularly suitable for high energy
calculations since the numerical evolution is provided for $Q^2$
in the range $10^{0 - 8}$ GeV$^2$ and for $x$ down to $10^{-6}$
($E_{\nu} \sim 10^{7}$ GeV). Values outside this range are
calculated by extrapolation.

\begin{figure}[h]
\begin{center}
\epsfig{figure=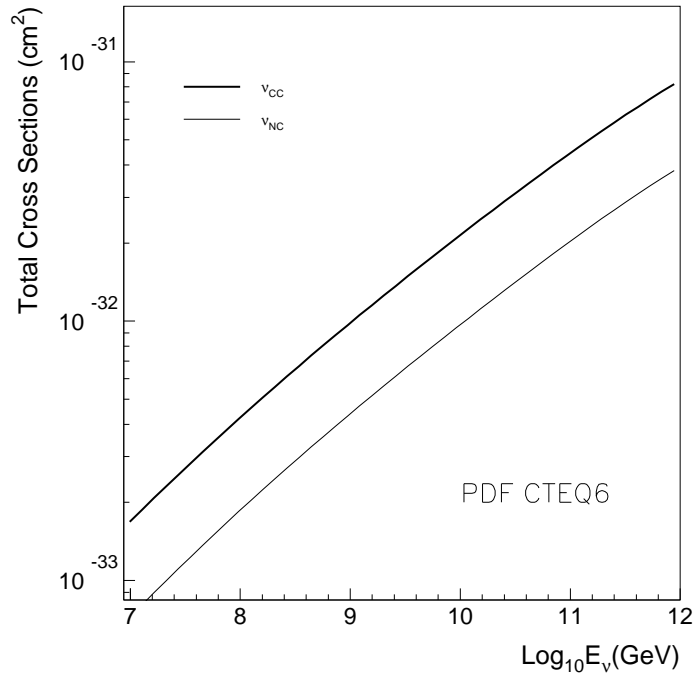,height=10cm} \caption{The $\nu_{\tau}$
total cross section for CC (thick line) and NC (thin line)
inelastic scattering off an isoscalar nucleon $N=(n+p)/2$ are here
reported. The calculation has been performed by using the CTEQ6
Parton Distribution Functions (PDF) \cite{cteq6}, according to the
prescription given in Ref. \cite{quiggcross}.} \label{crossneu}
\end{center}
\end{figure}
\begin{figure}[h]
\begin{center}
\epsfig{figure=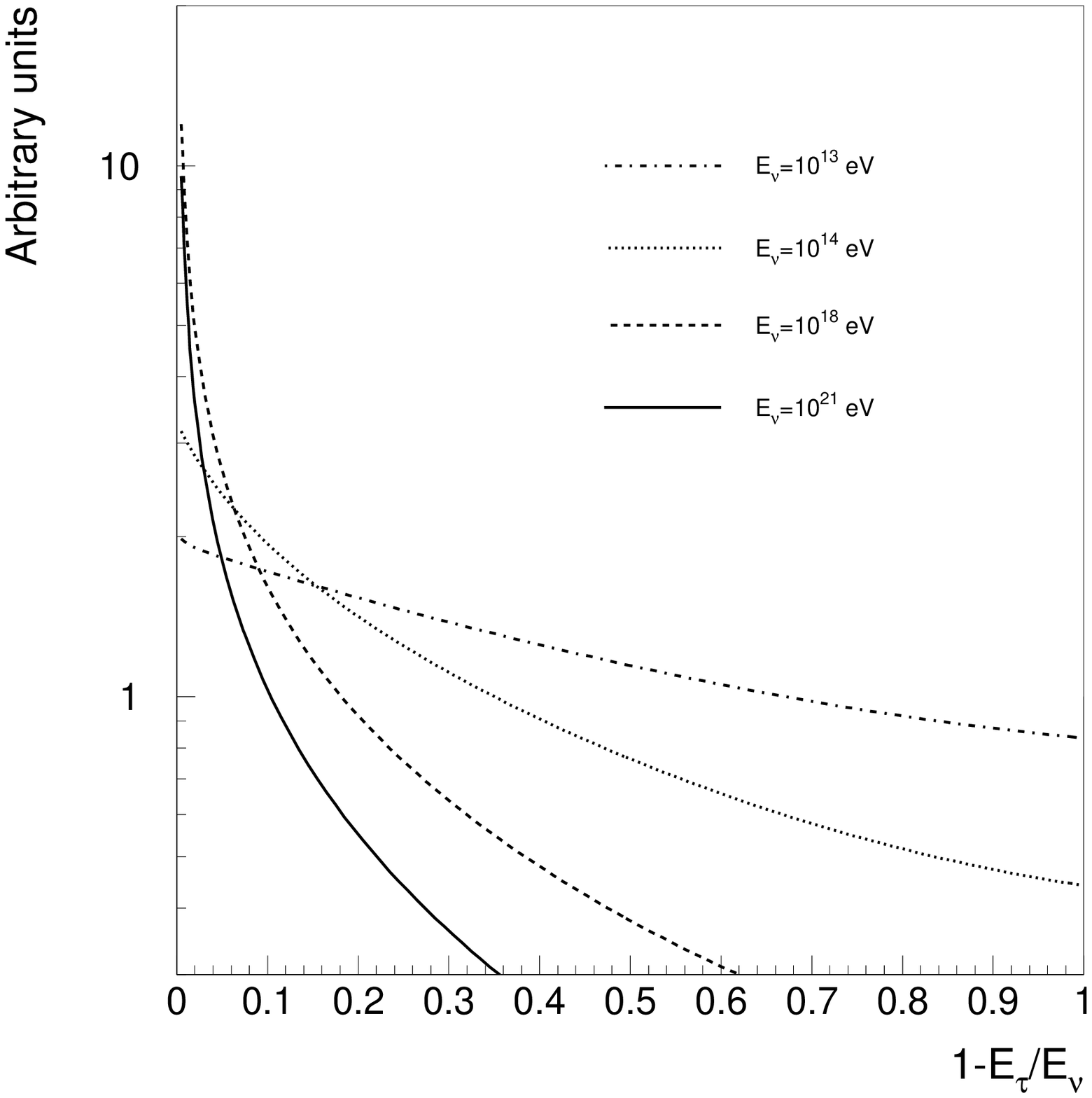,height=10cm} \caption{Distribution of the
inelasticity parameter, $y_{CC}=1 - E_{\tau}/E_{\nu}$, for
different neutrino energies. } \label{ydis}
\end{center}
\end{figure}

The total cross sections for $\nu_\tau$ CC and NC inelastic
scattering off an isoscalar nucleon, $N=(n+p)/2$ ($n=\#$ of
neutrons, $p=\#$ of protons), are shown in Figure~\ref{crossneu}.
The calculation for $\bar{\nu}_{\tau}$ is not shown because
antineutrino- and neutrino-nucleon cross sections become
indistinguishable for $E_{\nu}  > 10^{6}$ GeV. This is due to the
dominance at very high-energy, i.e small $x$, of the {\it sea}
quarks on the {\it valence} ones. Figure~\ref{ydis} shows the
distribution of the inelasticity parameter
$y_{CC}=1-E_{\tau}/E_{\nu}$, for different neutrino energies. An
average inelasticity $<y>$ can be defined by integrating the
distributions given in Figure~\ref{ydis}. Few relevant values of
$<y>$ are summarized in Table~\ref{table::y} for CC ($<y_{CC}>$)
and for NC ($<y_{NC}>$) interaction.

\begin{figure}[htb]
\begin{center}
\epsfig{figure=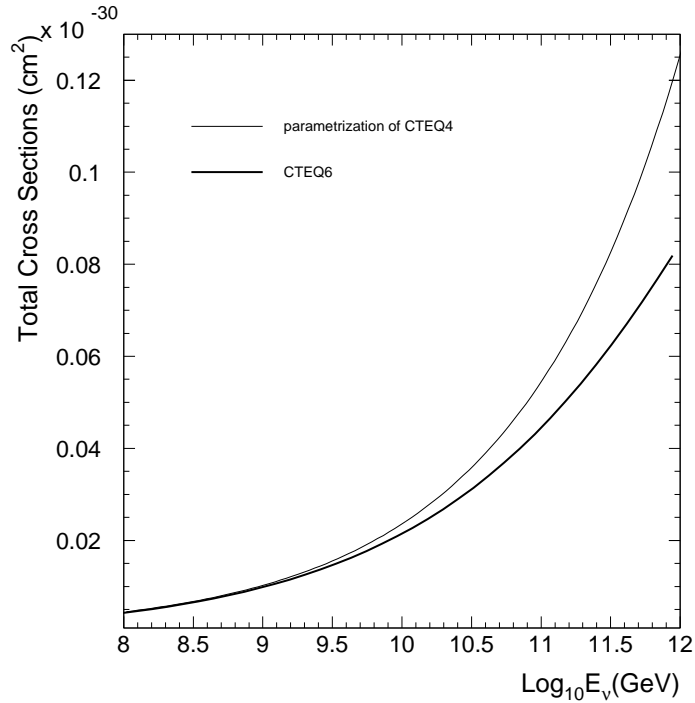,height=10cm} \caption{The total
$\nu_\tau$-nucleon CC cross sections based on both CTEQ4
\cite{quiggcross} and CTEQ6~\cite{cteq6} are reported.}
\label{comp}
\end{center}
\end{figure}

Figure~\ref{comp} shows a comparison between a CTEQ4-based
parametrization of CC cross section and the corresponding
calculation performed with CTEQ6. A substantial agreement is found
up to 10$^9$ GeV, whereas a discrepancy of at most 30\%  at
10$^{12}$ GeV is observed (CTEQ4 prediction being larger). In this
range of energy the uncertainty due to the lack of knowledge of
parton distribution functions is expected to be overwhelming.

\section{The $\tau$ energy losses}
\label{sec:tauloss}

Tau leptons with energy higher than 10$^{16}$ eV travelling in
matter may loose a consistent fraction of their energy before
decaying~\cite{dutta}. A precise knowledge of $\tau$ energy loss
is therefore required in order to draw reliable predictions of
expected signal rate at detectors. High-energy $\tau$'s
propagating through matter mainly interact by quasi-continuous
(ionization) and discrete energy loss mechanisms, mainly by
photonuclear interaction, direct electron-positron pair production
and Bremsstrahlung. The ionization energy loss dominates for
$\tau$--energy smaller than a few TeV, while radiative processes
become relevant at higher energies.

The direct electron pair production differential cross section has
been calculated by Kelner and Kotov in the framework of QED theory
\cite{kelner}. We have used the well-known parametrization
performed by Kokoulin and Petrukhin~\cite{koko}, which considers
the corrections for atomic and nuclear form factors. The
expression for the Bremsstrahlung differential cross section has
been derived by Andreev and Bugaev~\cite{Andreev}. It takes into
account the nuclear structure of the target (elastic and inelastic
form factors) and the exact contributions due to atomic electrons
(screening effect and Bremsstrahlung on electrons).

The complete formulas used here for computing electron pair
production and Bremss\-trahlung cross sections are the ones
reported in Ref. \cite{gmu}, with the only substitution of the
muon mass with the $\tau$ mass. Almost identical formulas are
given in Ref. \cite{Lohmann} and in Ref. \cite{dutta}, where a
slightly simplified formula for Bremsstrahlung (in agreement
within a few percent with the one of Ref.~\cite{gmu}) is actually
used. A complete list of $\tau$ matter cross sections written
within the same theoretical framework adopted here is also given
in Ref. \cite{bugaevtot}.

The photonuclear differential cross section is calculated
following the theoretical approach developed in Ref.
\cite{bugaev03}. According to this formalism, the cross section
for photonuclear interaction consists of two terms, where the
first one, obtained within the Vector Meson Dominance Model,
describes the non-perturbative contribution to the electromagnetic
structure functions. The parametrization given in Ref.
\cite{bugaev03} differs from the corresponding well known result
of Ref.~\cite{bugaev81} by few new terms, negligible for muons but
important for $\tau$'s. The second term, with parameters updated
with the most recent experimental data, describes the perturbative
QCD contribution, not negligible at extremely high energies
($E_{\tau}>10^{15}$ eV). The parametrization of these terms are
provided up to 10$^{9}$ GeV in Ref. \cite{sokalski}, whereas the
values at higher energy are obtained by extrapolation. In order to
derive the non-perturbative term we considered the recent
accelerator data coming from the experiments ZEUS and
H1~\cite{Zeus,H1}.

The average energy loss for a given discrete process $k$ can be
expressed in terms of the differential cross section,
$d\sigma^{k}/dv$, as follows
\begin{equation}
 -\Big\langle \frac{dE}{dx}\Big\rangle_{k}=\frac{N_{A}}{A} \, E_{\tau} \,
         \int_{v_{min}}^{v_{max}}\, v
 \frac{d\sigma^{k}}{dv}(v,E_{\tau}) \,dv \,= \beta_{k}(E_{\tau}) \, E_{\tau}\,\,\, ,
\label{losses}
\end{equation}
where $N_{A}$ is the Avogadro's number, $A$ is the mass number,
$v$ is the fraction of initial energy $E_{\tau}$ lost by the
$\tau$ at the occurrence of the process $k$ and $x$ is the
thickness of the crossed matter, expressed in g/cm$^{2}$.

\begin{figure}[htb]
\begin{center}
\epsfig{figure=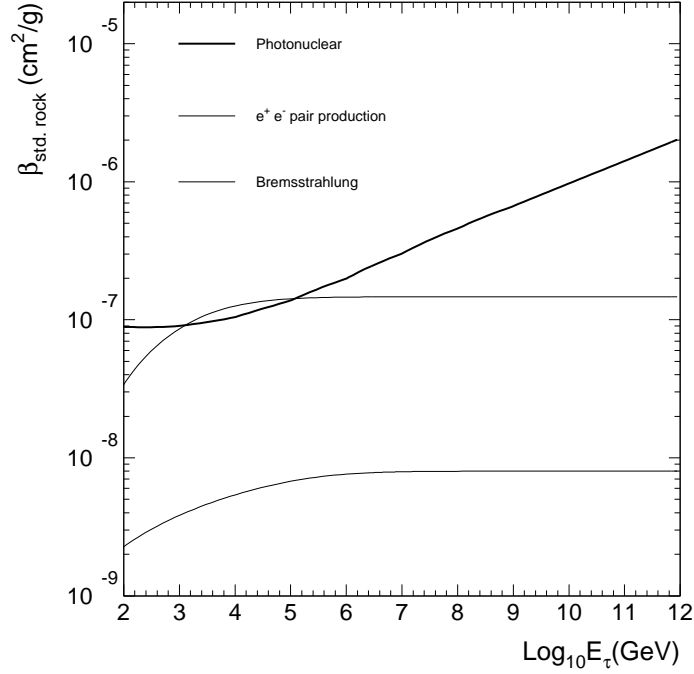,height=10cm} \caption{$\beta$ value for
photonuclear interaction (the most relevant process for $\tau$'s
at high energies), electron pair production and Bremsstrahlung.
Results are shown for standard rock (Z=11, A=22, $\varrho_s=2.65$
g/cm$^{3}$).} \label{beta}
\end{center}
\end{figure}

\begin{figure}[htb]
\begin{center}
\epsfig{figure=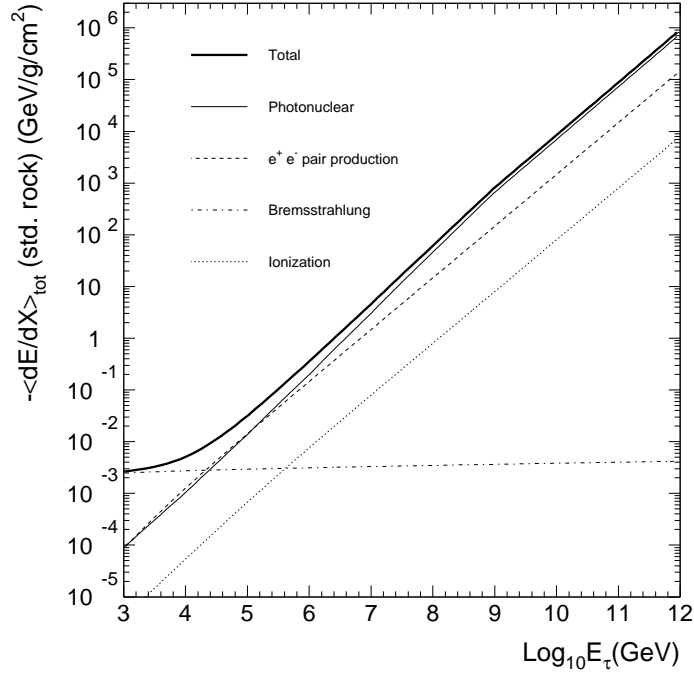,height=10cm} \caption{The total average
energy loss in standard rock is plotted (thick line). The
contributions of each individual electromagnetic process are also
reported (thin lines). Results are shown for standard rock (Z=11,
A=22, $\varrho_s=2.65$ g/cm$^{3}$).} \label{loss}
\end{center}
\end{figure}

\begin{figure}[htb]
\begin{center}
\epsfig{figure=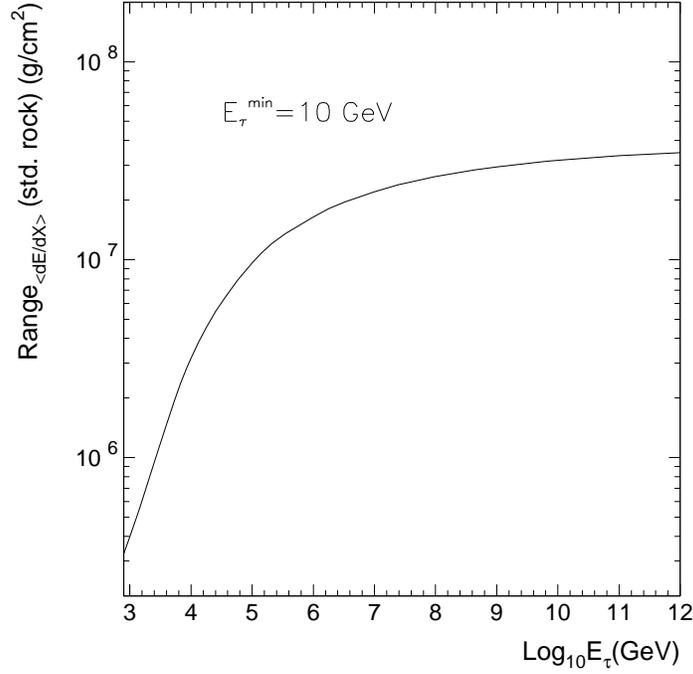,height=10cm} \caption{Range of average
energy loss as a function of energy $E_\tau$. $E_\tau^{min}$ has been fixed
at 10 GeV. Results are shown for standard rock (Z=11, A=22,
$\varrho_s=2.65$ g/cm$^{3}$).}
\label{range}
\end{center}
\end{figure}

Figure~\ref{beta} shows the $\beta$ values for photonuclear
interaction (the most relevant process at high energies), electron
pair production and Bremsstrahlung versus $\tau$ energy.
Figure~\ref{loss} shows the energy loss due to individual
electromagnetic processes and the total average energy loss
defined as
\begin{equation}
 -\Big\langle \frac{dE}{dx}\Big\rangle_{tot}=
 -\Big\langle \frac{dE}{dx}\Big\rangle_{ionization} +
\sum_{k} \beta_{k}E_{\tau}\,\,\, .
\end{equation}
Finally, Figure~\ref{range} shows the range of the average energy
loss,
\begin{equation}
\mathcal{R}(E_{\tau},E_{\tau}^{min})=\int_{E_{\tau}^{min}}^{E_{\tau}}
\,
       \frac{dE^{'}_{\tau}}{-\Big\langle dE/dx\Big\rangle_{tot}(E^{'}_{\tau})}
\,\,\,,
\label{range_eq}
\end{equation}
as a function of the initial, $E_{\tau}$, and the final,
$E_{\tau}^{min}$, $\tau$--energy. All results shown in
Figures~\ref{beta}~-~\ref{range} are calculated for standard rock
(Z=11, A=22, $\varrho_s=2.65$ g/cm$^{3}$). A detailed evaluation
of the average (effective) range would require a full stochastic
treatment of discrete interactions in order to correctly handle
fluctuations. As shown in Ref. \cite{lipari} and \cite{dutta},
fluctuations may strongly decrease the average range compared to
the range of the average energy loss. Results of a full simulation
of $\tau$ propagation through matter are also shown in Ref.
\cite{sokalski} and \cite{bottai}.


\section{The Earth--skimming events}
\label{sec:earthskimming}

\begin{figure}[htb]
\begin{center}
\epsfig{figure=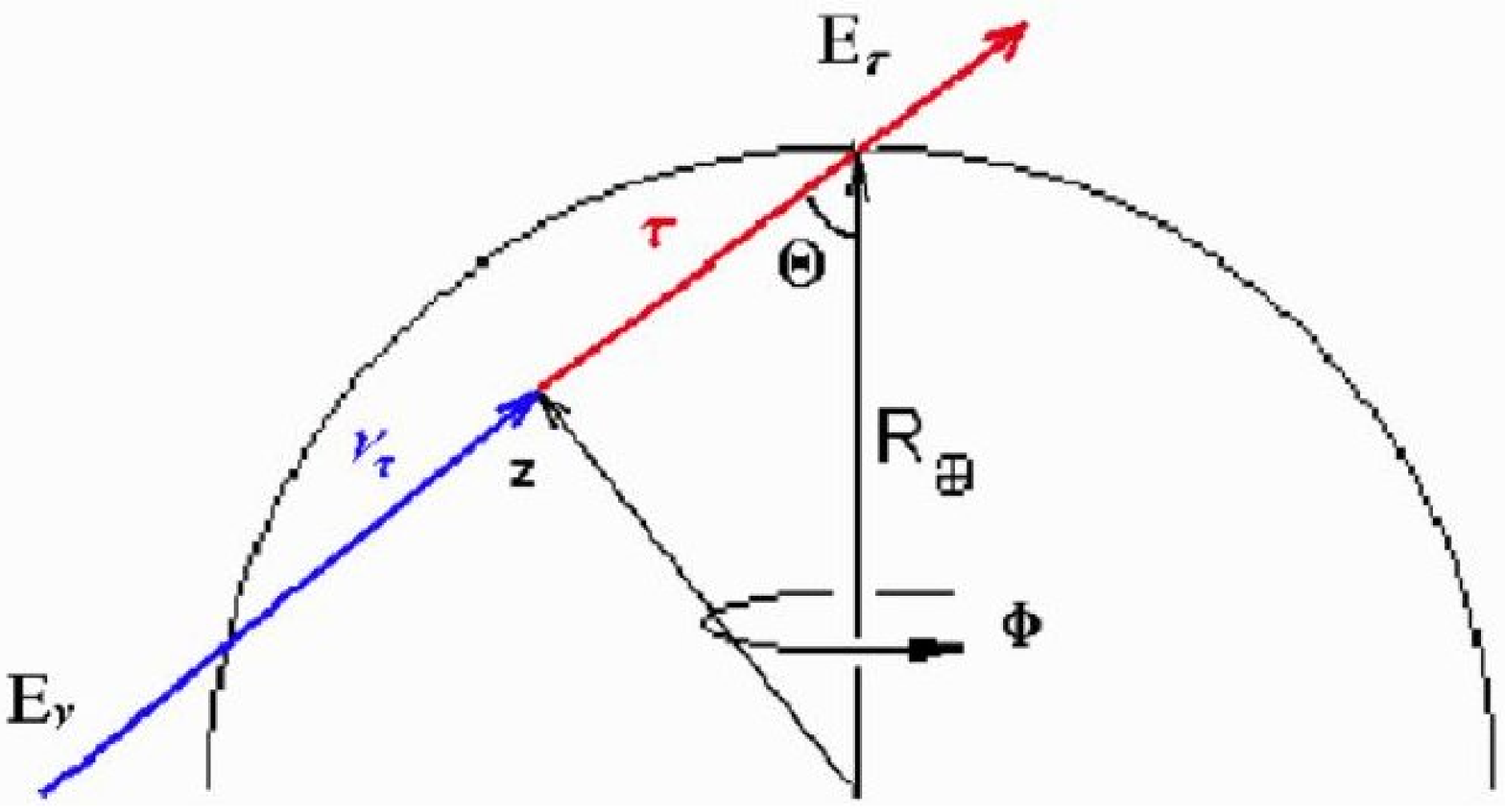,height=6cm} \caption{A neutrino
$\nu_\tau$ crosses the Earth with energy $E_\nu$ at a nadir angle
$\theta$ and azimuth angle $\phi$. Then it travels a distance $z$
before converting into a charged lepton $\tau$, which emerges from
the Earth surface with energy $E_\tau$ \cite{Feng:2001ue}.}
\label{geo1}
\end{center}
\end{figure}

Following the formalism developed in Ref. \cite{Feng:2001ue}, let
$\Phi_\nu$ be an isotropic flux of $\nu_\tau+\overline{\nu}_\tau$.
The differential flux of charged leptons emerging from the Earth
surface with energy $E_\tau$ is given by
\begin{equation}
\frac{d\Phi_\tau(E_\tau,\theta,\phi)}{dE_\tau\,d\Omega}=
\int dE_\nu \, \frac{d\Phi_\nu(E_\nu,\theta,\phi)}{dE_\nu\,d\Omega}
\, K(E_\nu,\,\theta;\,E_\tau)\,\,\, ,
\label{eq:1}
\end{equation}
where $K(E_\nu,\,\theta;\,E_\tau)$ is the probability that an
incoming neutrino crossing the Earth with energy $E_\nu$ and nadir
angle $\theta$ produces a lepton emerging with energy $E_\tau$
(see Figure~\ref{geo1}). In Eq.(\ref{eq:1}), due to the very high
energy of $\nu_\tau$, we can assume that in the process $\nu_\tau
\, + \, N \rightarrow \tau \, + \, X$ the charged lepton is
produced along the neutrino direction.
This process can occur if and only if the following conditions are
fulfilled:\\
a) the $\nu_\tau$ with energy $E_\nu$ has to survive along a
distance $z$ through the Earth;\\
b) the neutrino converts into a $\tau$ in the
interval $z, z+dz$;\\
c) the created lepton emerges from the Earth before decaying.\\

\noindent a) The probability $P_a$ that a neutrino with energy
$E_\nu$ crossing the Earth survives up a certain distance $z$ is
\begin{equation}
P_a=\exp\left\{-\int_0^z\,\frac{dz'}{\lambda_{CC}^\nu(E_\nu,\,\theta,\,z')}\right\}\,\,\,,
\label{eq:2}
\end{equation}
where
\begin{equation}
\lambda_{CC}^\nu(E_\nu,\,\theta,\,z)=
\frac{1}{\sigma_{CC}^{\nu N}(E_\nu)\,\varrho[r(\theta,\,z)]\,N_A}
\label{eq:3}
\end{equation}
is the CC interaction length in the rock, $\varrho[r(\theta,\,z)]$
is the Earth's density at distance $r$. The distance $r$ is given
by $r^2(\theta,\,z) = R_\oplus^2+z^2-2{R_\oplus}z \cos\theta$,
where $R_\oplus \simeq 6370\,\mathrm{km}$ is the average Earth
radius. The expression of $P_a$ does not take into account the
atmosphere crossed by the $\nu_\tau$ before entering the Earth
surface, because the CC interaction length in the air is almost
three orders of magnitude larger than in the rock. \\

\noindent
b) The probability for $\nu_\tau \rightarrow \tau$  conversion in the
interval $[z,\,z+dz]$ is
\begin{equation}
P_b = \frac{dz}{\lambda_{CC}^\nu\,(E_\nu,\,\theta;\,z)}\,\,\, .
\label{eq:4}
\end{equation}
Here a comment is in turn. In order to produce a $\tau$ emerging
with enough energy to generate a electromagnetic shower detectable
by FD,  the charged lepton propagation in the rock is limited.
Moreover, since an EeV neutrino has a $\lambda_{CC}^\nu \sim 500$
km, only quite horizontal $\nu_\tau$ will be able to produce
detectable events. The charged current interaction will then take
place near to the Earth surface where the average density is
almost constant and equal to $\varrho_s \simeq 2.65$ g/cm$^{3}$.\\

\noindent c) The probability $P_c$ that a charged lepton loosing
energy survives as it travels through the Earth is described by
the coupled differential equations:
\begin{eqnarray}
\frac{dP_c}{dz}&=&-\frac{m_\tau}{c \, \tau_\tau\,E_\tau}\,P_c\,\,\, ,
\label{eq:7}\\
\frac{dE_\tau}{dz}&=&- \left(\beta_\tau
+\gamma_\tau\,E_\tau\right) \,E_\tau \, \varrho_s \,\,\, .
\label{eq:8}
\end{eqnarray}
Here $m_\tau = 1.8 {\cdot} 10^9$ eV, $\tau_\tau \simeq 3.4 {\cdot}
10^{-13}\,$s denotes the $\tau$ mean lifetime, whereas the
parameters $\beta_\tau  \simeq 0.71 {\cdot} 10^{-6}$ cm$^2$
g$^{-1}$ and $\gamma_\tau \simeq 0.35 {\cdot} 10^{-18}$ cm$^2$
g$^{-1}$ GeV$^{-1}$, as discussed in Section \ref{sec:tauloss},
fairly describe the $\tau$ energy loss in matter. The set of
equations (\ref{eq:7}), (\ref{eq:8}) can be solved by observing
that, following the results presented in section \ref{sec:nucross}
and shown in Table \ref{table::y}, the tau lepton produced at $z$
carries an average energy which is a function of $E_\nu$. Let
$E^0_{\tau} = E^0_{\tau}(E_\nu)=(1-<y_{CC}>)E_\nu$ be the
transferred energy. By solving Eq.s (\ref{eq:7}) and (\ref{eq:8})
at the emerging point on the Earth surface one has:
\begin{eqnarray}
P_c &=&  \left( F(E_\nu,E_\tau)\right)
^{\omega} \,
\exp\left\{-\frac{m_\tau}{c \tau_\tau \beta_\tau \varrho_s}\left(\frac{1}{E_\tau}-
\frac{1}{E_\tau^0(E_\nu)}\right)\right\}\,\,\, , \label{eq:9}
\\
E_\tau & = & \frac{\beta_\tau \, E^0_{\tau}(E_\nu)
\,
\exp\left\{-
\varrho_s \, \beta_\tau (2 R_\oplus \cos\theta- z)\right\}}
{\beta_\tau + \gamma_\tau \, E^0_{\tau}(E_\nu) \left(1-\exp\left\{-
\varrho_s
\,
\beta_\tau (2 R_\oplus \cos\theta- z)\right\}\right)}\,\,\, ,
\label{eq:10}
\end{eqnarray}
where
\begin{equation}
F(E_\nu,E_\tau) \equiv
\frac{E^0_{\tau}(E_\nu)(\beta_\tau +
\gamma_\tau E_\tau) }{E_\tau(\beta_\tau +\gamma_\tau \, E^0_{\tau}(E_\nu)
)}\,\,\,,
\,\,\,\,\,\,\,\,\,
\omega \equiv \frac{m_\tau \, \gamma_\tau}{c \tau_\tau \beta_\tau^2 \varrho_s}\,\,\,
.
\label{eq:10a}
\end{equation}
The above results improve the ones obtained in Ref.
\cite{Feng:2001ue}, where a simpler parametrization for
the $\tau$ energy loss was adopted.\\

The energy $E_\tau$ of the exiting lepton must be consistent with
Eq.(\ref{eq:10}). This condition is enforced by the
$\delta$-function:
\begin{equation}
P_d=\delta\left(E_\tau-\frac{\beta_\tau \, E^0_{\tau}(E_\nu)
\,
\exp\left\{-
\varrho_s \, \beta_\tau (2 R_\oplus \cos\theta- z)\right\}}
{\beta_\tau + \gamma_\tau \, E^0_{\tau}(E_\nu) \left(1-\exp\left\{-
\varrho_s
\,
\beta_\tau (2 R_\oplus \cos\theta- z)\right\}\right)}\right)\,\,\, .
\label{eq:11}
\end{equation}
By using the expressions for the different probabilities
(\ref{eq:2}), (\ref{eq:4}), (\ref{eq:9}), and (\ref{eq:11}) the
kernel reads
\begin{equation}
K(E_\nu,\,\theta;\,E_\tau)=\int_0^{2R_\oplus \cos\theta}\,P_a \, P_b \, P_c
\, P_d \,\,dz\,\,\, .
\label{eq:12}
\end{equation}
Once the integration over $z$ is performed, one gets the simple
result
\begin{eqnarray}
&&K(E_\nu,\,\theta;\,E_\tau)=
\frac{\sigma_{CC}^{\nu N}(E_\nu) \, N_A}{E_\tau (\beta_\tau+\gamma_\tau \, E_\tau)} \,
\left(F(E_\nu,E_\tau)\right)^{\xi} \,
\nonumber\\
&&{\times}
\exp\left\{-\frac{m_\tau}{c \tau_\tau \beta_\tau \varrho_s}\left(\frac{1}{E_\tau}-
\frac{1}{E_\tau^0(E_\nu)}\right)-2 R_\oplus
\cos{\theta} \, \sigma_{CC}^{\nu N}(E_\nu) \, \varrho_s \, N_A \right\}\,\,\, ,
\label{eq:13}
\end{eqnarray}
where
\begin{equation}
\xi \equiv
\left(\omega
+ \frac{\sigma_{CC}^{\nu N}(E_\nu) \, N_A}{\beta_\tau} \right)\,\,\, .
\label{eq:14}
\end{equation}
Eq.(\ref{eq:11}) requires that two conditions are fulfilled. The
first one is $E^0_{\tau}(E_\nu) \geq E_{\tau}$ (obviously
verified), while the second is
\begin{equation}
\cos\theta \geq \cos\theta_{min} =\frac{1}{ 2 \,R_\oplus \, \beta_\tau\, \varrho_s } \,
\log\left( F(E_\nu,E_\tau)\right)\,\,\, .
\label{eq:15}
\end{equation}
The expression (\ref{eq:13}) for the kernel
$K(E_\nu,\,\theta;\,E_\tau)$ leads to the total rate of upgoing
$\tau$'s showering on the Auger detector, and thus {\it
potentially} detectable by the FD:
\begin{eqnarray}
\frac{d\textsl{N}_\tau}{dt}&=&2 \pi S \, D \,\int_{E_\nu^{min}}^{E_\nu^{max} } dE_\nu
\int_{E_\tau^{th}}^{E_\tau^0(E_\nu)} dE_\tau \int_{\cos\theta_{min}}^{1}\,
\frac{d\Phi_\nu(E_\nu)}{dE_\nu\,d\Omega} \,  \nonumber\\
&{\times}&
 K(E_\nu,\,\theta;\,E_\tau) \, \left(1-\exp\left\{-\frac{H \, m_\tau}
{c\tau_\tau \, E_\tau}\right\}\right)
\, \varepsilon \, \cos\theta \,d(\cos\theta)  \,\,\,,
\label{eq:16}
\end{eqnarray}
where we have used the isotropy of the considered neutrino flux.
In Eq.(\ref{eq:16}) the quantity $S=3000\,\mathrm{km}^2$ is the
geometrical area covered by the Auger apparatus, $D\sim 10\% $ is
the duty cycle for fluorescence detection, $ E_\tau^{th} \simeq
10^{18}$ eV is the energy threshold for the fluorescence process,
and $E_\nu^{min}$ is the minimum neutrino energy capable of
producing a $\tau$ at detection threshold. The quantity
$E_\nu^{max}$ is the endpoint of the neutrino flux.

The exponential term in the r.h.s. of Eq.(\ref{eq:16}) accounts
for the decay probability of a $\tau$ (showering probability) in a
distance $H$ from the emerging point on the Earth surface. A
detailed calculation of the fraction of Earth--skimming $\tau$'s
decaying inside the fiducial volume would require a full Monte
Carlo simulation, which is not the aim of the present analysis. A
reasonable way to estimate such a fraction is to require that
$\tau$ leptons travel a distance less than $H$ before decaying. We
choose for simplicity $H=30$ km, which is the radius of an
emisphere approximatively containing the entire PAO apparatus.
However, this ansatz provides a conservative estimate of the
number of events. FD detection efficiency $\varepsilon$ has been
taken by Ref. \cite{Auger}.

In Eq.(\ref{eq:16}) the integration over $\cos\theta$ can be
easily performed and this yields to
\begin{eqnarray}
\frac{d\textsl{N}_\tau}{dt}= D \, \int_{E_\nu^{min}}^{E_\nu^{max} } dE_\nu \,
\frac{d\Phi_\nu(E_\nu)}{dE_\nu\,d\Omega} \, A(E_\nu)\,\,\, ,
\label{eq:17}
\end{eqnarray}
where the {\it effective aperture} of the apparatus is here defined as
\begin{eqnarray}
&&A(E_\nu)=\frac{\pi S}{2 \, R_\oplus^2 \, N_A \,
\varrho_s^2} \,
\int_{E_\tau^{th}}^{E_\tau^0(E_\nu)} dE_\tau \,
\frac{\left(F(E_\nu,E_\tau)\right)^{\omega}}{E_\tau (\beta_\tau+\gamma_\tau \, E_\tau)} \,
\nonumber\\
&&{\times} \exp\left\{-\frac{m_\tau}{c \tau_\tau \beta_\tau
\varrho_s}\left(\frac{1}{E_\tau}-
\frac{1}{E_\tau^0(E_\nu)}\right)\right\}\,
 \left(1-\exp\left\{-\frac{H \, m_\tau}
{c\tau_\tau \, E_\tau}\right\}\right)
\, \varepsilon \nonumber\\
&&{\times} \frac{1}{\sigma_{CC}^{\nu N}} \, \left[ \left(1 +
\frac{\sigma_{CC}^{\nu N}\, N_A}{\beta_\tau}
\log\left( F(E_\nu,E_\tau)\right)\right)  \right. \nonumber\\ &&-
\left. \left( 1 + 2
\, R_\oplus \,
\sigma_{CC}^{\nu N} \, \varrho_s  \, N_A\right) \exp\left\{ - 2 \, R_\oplus \,
\sigma_{CC}^{\nu N} \, \varrho_s  \, N_A\right\} \, \left(F(E_\nu,E_\tau)\right)^
{\sigma_{CC}^{\nu N} N_A/\beta_\tau }
  \right].\,\,\,\,\,
\label{eq:aperture}
\end{eqnarray}
\begin{figure}
\begin{center}
\epsfig{figure=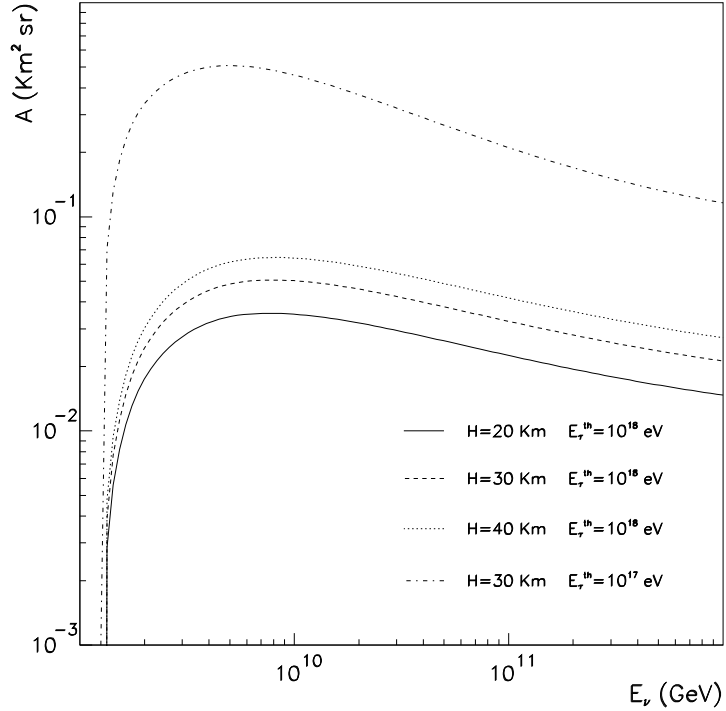,height=10cm} \caption{The {\it
effective aperture}, $A(E_\nu)$, as defined in
Eq.(\ref{eq:aperture}) is here plotted for $E^{th}_\tau=10^{17}$
eV and $H=30$ km, and for $E^{th}_\tau=10^{18}$ eV with $H=20$,
30, and 40 km, respectively.} \label{fig-aperture}
\end{center}
\end{figure}
In Figure \ref{fig-aperture} the quantity $A(E_\nu)$ is plotted
versus the neutrino energy for $E^{th}_\tau=10^{17}$ eV and $H=30$
km, and for $E^{th}_\tau=10^{18}$ eV with $H=20$, 30, and 40 km,
respectively. The aperture for $E^{th}_\tau=10^{18}$ eV shows a
maximum near $E_\nu \sim 10^{19} \,$ eV, sensibly dependent on the
parameter $H$. For a lower energy threshold, for example $10^{17}$
eV, $A(E_\nu)$ would scale as shown in Figure \ref{fig-aperture}.

A similar analysis for the PAO Surface Detector has been performed
via a Monte Carlo simulation in Ref. \cite{Bertou:2001vm}. In
particular, one could compare the aperture given in Figure
\ref{fig-aperture} with Figure 9 of Ref. \cite{Bertou:2001vm}. A
direct comparison of the two calculations is difficult to be
performed since the detector efficiency and the energy threshold
of FD and SD differ. In our analysis we use an energy threshold
$E^{th}_\tau=10^{18}$ eV. However, $A(E_\nu)$ approaches the
aperture of Ref. \cite{Bertou:2001vm} as $E^{th}_\tau$ decreases.

In Table \ref{table::events} the number of $\tau$-shower
Earth--skimming events expected per year at the FD detector is
reported for the models of neutrino flux discussed in section
\ref{sec:nuflux}.

\begin{table}[thb]
\centering
\begin{tabular}{||c|c|c|c|c|c||}
\hline\hline
$dN_\tau/dt$ at FD & GZK-WB & GZK-L & GZK-H & TD & NH\\
\hline\hline
\# of UP events/year & 0.02 & 0.04 &  0.09 & 0.11 &  0.25 \\
\hline
\end{tabular}
\caption{Yearly rate of upgoing (UP) events at FD for different
neutrino models given in Figures \ref{nu_gzk} and
\ref{nu_exotic}.} \label{table::events}
\end{table}

In Ref. \cite{Feng:2001ue} an explicit plot of the aperture as
defined in our Eq.(\ref{eq:aperture}) is not given. However, it is
worth observing that the number of events reported in Table
\ref{table::events} for TD model is in fair agreement with the
corresponding result of Ref. \cite{Feng:2001ue}, once properly
normalized to PAO effective area. Other cases are not immediately
comparable since fluxes are different.
\begin{figure}
\begin{center}
\epsfig{figure=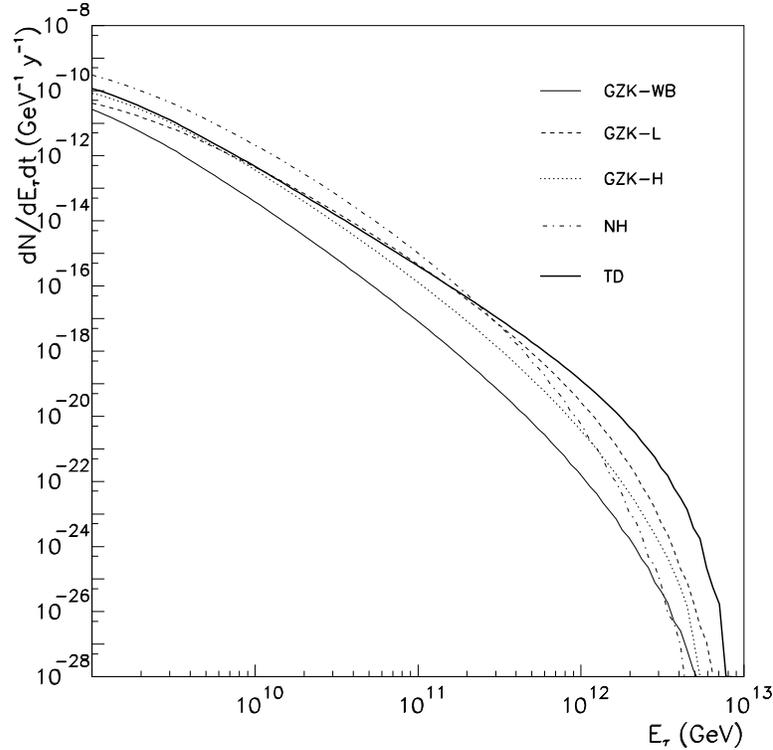,height=10cm}
\caption{The quantity $dN_\tau/dE_\tau \, dt$
is plotted for the different neutrino models reported in Figures
\ref{nu_gzk} and \ref{nu_exotic}.}
\label{fig-kernel}
\end{center}
\end{figure}

In Figure \ref{fig-kernel} the energy spectra of Earth--skimming
events at FD are shown for the neutrino models of Figures
\ref{nu_gzk} and \ref{nu_exotic}. As expected, the maximum number
of events is reached near to the assumed FD threshold ($\simeq$ 1
EeV), even though the maxima of neutrino fluxes are at higher
energy. This can be easily understood by observing that tau
leptons emerging inside the Auger surface with large energy and
almost horizontally will probably decay far from the apparatus and
are thus undetectable by the FD.
\begin{figure}
\begin{center}
\epsfig{figure=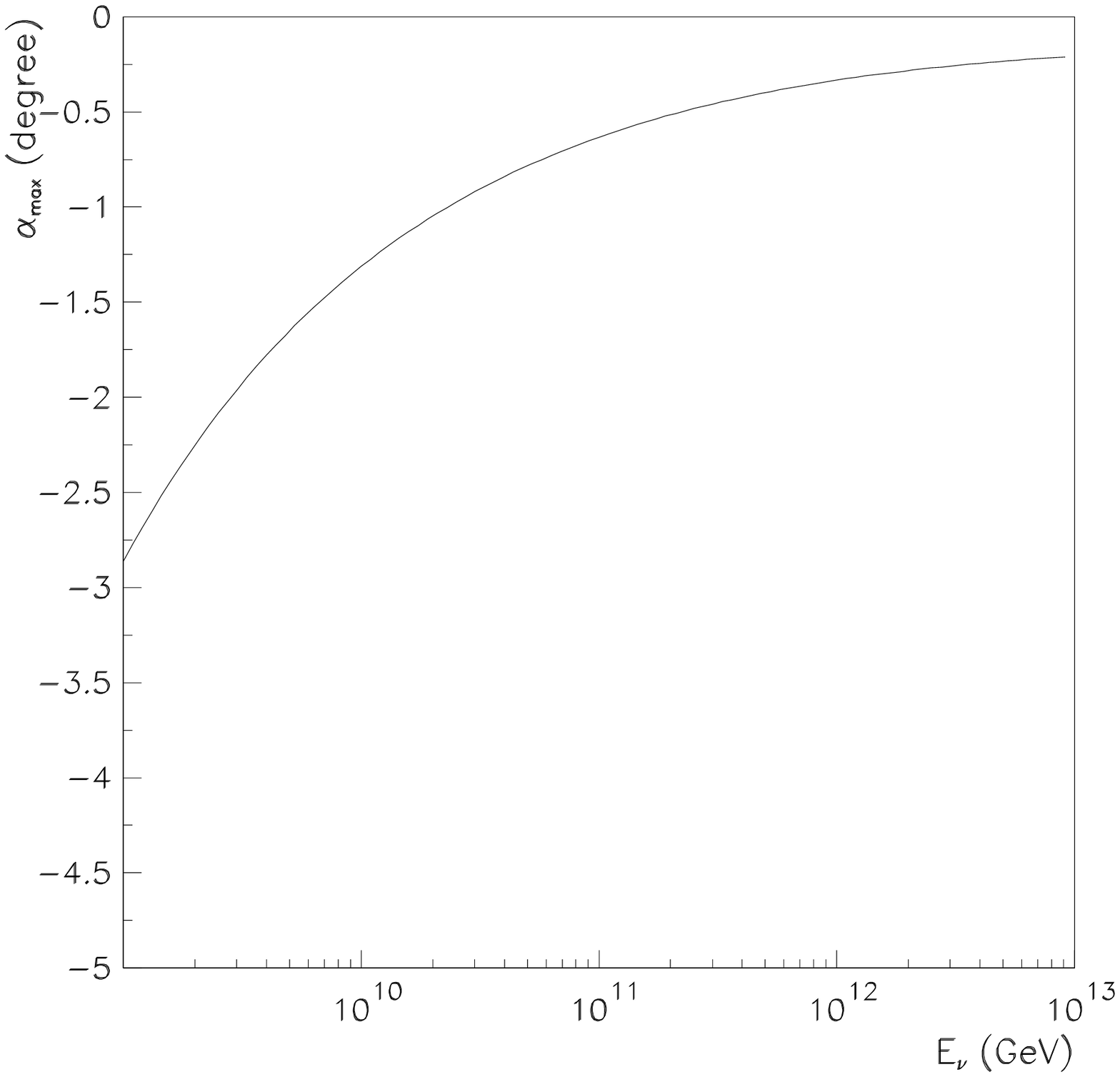,height=10cm} \caption{The most likely
exit angle with respect to the horizontal for the emerging $\tau$
is plotted versus the energy of the primary neutrino.}
\label{alphamax}
\end{center}
\end{figure}
By using the integrand in the r.h.s. of Eq.(\ref{eq:17}),  the
angle with respect to the horizontal of the Earth-skimming event
$\alpha_{max} = \theta_{max} - \pi/2$ can be determined. Here,
$\theta_{max}$ denotes the nadir angle for which the kernel of
Eq.(\ref{eq:17}) has the maximum. This quantity is a function of
neutrino energy via $\sigma_{CC}^{\nu N}$ and is shown in Figure
\ref{alphamax}.

\begin{figure}
\begin{center}
\epsfig{figure=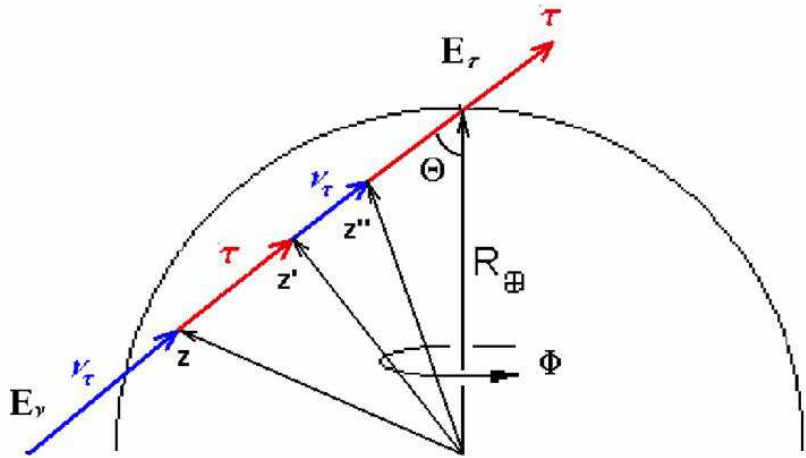,height=8cm}
\caption{Second order regeneration process
for $\nu_\tau$.}
\label{geo2}
\end{center}
\end{figure}

As pointed out for example in Ref. \cite{Beacom:2001xn},
$\nu_\tau$ crossing deeply the Earth could experience a {\it
regeneration phenomenon} which would eventually let $\tau$'s
propagate up to the detector(see Figure \ref{geo2}). Furthermore,
this process would also lead to the production at the CC vertex of
secondary electron or muon neutrinos, then decaying in electrons
or muons.

The contribution of regenerated $\nu_\tau$'s to the
Earth--skimming event rate is negligible, since the process shown
in Figure \ref{geo2} is a second order transition in the weak
coupling constant. Moreover, for $E_{\tau}^{th} \geq 5 \cdot
10^{17}$ eV the Earth strongly suppresses the expected signal due
to $\tau$ energy losses \cite{Becattini:2000fj, Bottai:2002nn}. As
shown in Ref. \cite{Becattini:2000fj, Bottai:2002nn}, the value of
$E_{\tau}^{th}$ plays a crucial role in the estimate of the
regenerated neutrino event rate, which results negligible for
threshold energies larger than $10^{18}$ eV. Higher order
processes including NC interactions, would also give negligible
modifications to the numbers presented in Table
\ref{table::events}. Finally, the associated secondary electron
and muon neutrino flux does not enhance the Earth--skimming signal
at PAO, since the charged leptons they produce are either absorbed
or not detectable by FD.

As well known, neutrino induced extensive air showers (see Ref.
\cite{Ambrosio:2003nr} for an extended discussion) can be
disentangled by the ordinary cosmic ray background only for very
inclined and deep showers. An estimate of the expected number of
downgoing events (DW) within 30 km from the FD detector is given
in Table \ref{table::number} for different fluxes and increasing
zenith angle. The DW event rates result to be comparable with the
Earth--skimming ones of Table \ref{table::events}. As a final
remark, the DW and Earth--skimming event rates are both affected
by the uncertainty on the $\nu$-nucleon cross section, but, unlike
the DW events, the Earth-skimming rate is also affected by the
uncertainty on the tau energy losses in rock.

\begin{table}[thb]
\centering
\begin{tabular}{||c|c|c|c|c|c||}
\hline\hline
 & GZK-WB & GZK-L & GZK-H & TD & NH\\
\hline\hline
\# of DW events/year ($\geq 60^\circ$)& $0.04$ & $0.08$ &  $0.22$ & $0.25$ &  $0.54$ \\
\# of DW events/year ($\geq 70^\circ$)& $0.02$ & $0.04$ &  $0.10$ & $0.12$ &  $0.25$ \\
\# of DW events/year ($\geq 80^\circ$)& $0.005$ & $0.01$ &  $0.03$ & $0.03$ &  $0.07$ \\
\hline
\end{tabular}
\caption{In the table are reported the expected number of
downgoing events (DW) within 30 km from the FD detector for
different fluxes and increasing zenith angle.}
\label{table::number}
\end{table}

\section{Conclusions}
\label{sec:conclusions}

Ultra High Energy $\nu_\tau$'s ($E_\nu \ge 1 \, EeV$) could have
real detection chance at giant surface apparatus like PAO. Almost
horizontal tau neutrinos crossing distance in the Earth of the
order of their interaction length might produce $\tau$-shower
Earth--skimming events potentially detectable by the Auger FD.

In this paper for a representative sample of $\nu_\tau$ fluxes
either produced as {\it cosmogenic neutrinos} or in more exotic
scenarios, the number of $\tau$-shower Earth--skimming events per
year expected at the FD of the Pierre Auger Observatory has been
computed. For this calculation we have used the CTEQ6~\cite{cteq6}
parton distribution functions in the DIS factorization scheme and
recent estimate of radiative $\tau$ energy losses in matter. A
decrease of $\sigma_{CC}^{\nu N}$ allows $\tau$ to emerge with a
smaller nadir angle (less horizontal) and thus increases the
number of Earth--skimming events. In the relevant region of
$E_\nu$, we essentially have $dN_\tau/dt \propto (\sigma_{CC}^{\nu
N})^{-1}$; thus a factor $1/2$ in the neutrino-nucleon cross
section leads to an almost double number of events at FD. As a
further remark, we point out that the level of theoretical
uncertainties on cross-sections at extremely high energy is very
large (like in case of new physics above the TeV energy region),
due to the poor experimental knowledge of parton density functions
for very small $x$.

The number of events per year at the FD of the Pierre Auger
Observatory is presented in Table \ref{table::events} for the
different neutrino models reported in Figure \ref{nu_gzk} and
\ref{nu_exotic}. For a five years PAO detection campaign (South
plus North) we essentially have to scale these numbers by a factor
ten and thus at least the exotic models could produce detectable
events. However, it is worth noticing that this estimate has been
performed under a rather conservative assumption on the
possibility of $\tau$ showering on the apparatus: we have used an
average value for the parameter $H$ which essentially allows only
one particle out of three to decay inside the detection fiducial
volume. In a paper in progress a more sophisticated simulation is
going to be performed by taking into account the real morphology
of the PAO site. This analysis should slightly increase the
numbers presented in Table \ref{table::events}, which however give
a reasonable estimate of the number of events at FD.

\section*{Acknowledgments}

We would like to thank  I. A. Sokalski and E. Bugaev for the extremely
helpful suggestions on the matter of $\tau$ photonuclear interaction. The
authors would also like to thank M. Ambrosio, F. Guarino and S. Pastor for
valuable comments. Work of D.S. supported in part by NASA ATP grant
NAG5-13399.

\end{document}